\documentclass[a4paper,prd,twocolumn,showpacs,amsmath,amssymb]{revtex4}

\usepackage{dcolumn}   
\usepackage{bm}        
\usepackage{graphicx}  

\def\ba{\begin{eqnarray}}
\def\ea{\end{eqnarray}}
\def\be{\begin{equation}}
\def\ee{\end{equation}}
\def\F{\mathcal{F}}

\def\etal{{\it et al.}}

\begin{document}


\title{A decision between Bayesian and Frequentist upper limit in
  analyzing continuous Gravitational Waves} 
\author{Iraj Gholami}\thanks{Email: iraj.gholami@theorie.physik.uni-goettingen.de}

\affiliation{Georg-August-Universit\"at G\"ottingen,~Institut f\"ur
  Theoretische Physik, Friedrich-Hund-Platz 1, D-37077 G\"ottingen,
  Germany}
\affiliation{Max-Plank-Institut f\"ur Gravitationsphysik,~(Albert
  Einstein Institute), Am M\"uhlenberg 1, 14476 Potsdam, Germany}


\begin{abstract}

Given the sensitivity of current ground-based Gravitational Wave (GW)
detectors, any continuous-wave signal we can realistically expect will
be at a level or below the background noise.  Hence, any data analysis
of detector data will need to rely on statistical techniques to
separate the signal from the noise. While with the current sensitivity
of our detectors we do not expect to detect any true GW signals in our
data, we can still set upper limits (UL) on their amplitude. These upper
limits, in fact, tell us how weak a signal strength we would
detect. In setting upper limit using two popular method, Bayesian and
Frequentist, there is always the question of a realistic results. In
this paper, we try to give an estimate of how realistically we can set
the upper limit using the above mentioned methods. And if any, which one
is preferred for our future data analysis work.
\end{abstract}


\maketitle

\section{Introduction}

Gravitational Waves (GWs), ripples in space-time which travel at the
speed of light, are a fundamental consequence of Einstein's General
Theory of Relativity. Due to the great distance to any likely
detectable sources of GWs, the signal amplitude reaching us will be
very small. Because of limitations in technology, there have been no
direct detections of GWs so far. However, with the future generation
of detectors, like Advanced LIGO, we should be able to detect a
variety of sources.

Because of their nature, continuous GWs (emitting from axisymetric
rotating neutron stars) reaching Earth are expected to be extremely
weak. Therefore even with the quite significant sensitivity of our
current detectors, it will be difficult to detect them. One
possible way to increase the overall signal compared to the background
noise (signal-to-noise ratio (SNR)) is to coherently integrate the
data for \textit{several days} up to \textit{few years}.

The basic problem in GW detection is to identify a gravitational
waveform in a noisy background. Because all data streams contain
random noise, the data are just a series of random values and
therefore the detection of a signal is always a decision based on
probabilities. The aim of detection theory is therefore to assess this
probability.

The basic idea behind the current methods of signal detection is that
the presence of a signal will change the statistical characterization
of the data $x(t)$, in particular its probability distribution
function (pdf) $P(x)$. Recall that the pdf is defined so that the
probability of a random variable $x_i$ lies in an interval between
$x(t)$ and $x(t)+dx$ is $P(x)dx$.  Let us denote by $P(x|0)$ the
probability of a random process $x(t)$ (representing our data) in the
absence of any signal, and by $P(x|h)$ the probability of that same
process when a signal $h(t)$ is present.  Given a particular
measurement $x(t)$ obtained with our detector, is its probability
distribution given by $P(x|0)$ or $P(x|h)$?  In order to make that
decision, we need to make a rule called a \textit{statistical test}.

There are several approaches to find an appropriate test, notably
the Bayesian, Minimax and Neyman-Pearson approach (for an overview, we
refer the reader to Jaranowski and ~Kr\'olak, \cite{LRR-JK} and the
references listed therein). In the end, however, these three
approaches lead to the same test, namely the \textit{likelihood ratio
test} \cite{LRR-JK, davis}.

Among the three main approaches, the Neyman-Pearson approach is often
used in the detection of gravitational waves
\cite{schutz-gr-qc-9710080}. This approach is based on maximizing the
detection probability (equivalently minimizing the false dismissal
rate) for fixed false alarm rate, where the detection probability is
the probability that the random value of a process which contains the
signal will pass our test, while the false alarm probability is the
probability that data containing no signal will pass the test
nonetheless. Mathematically, we can express the Detection and False
Alarm probabilities as
\cite{schutz-gr-qc-9710080}
\ba
P_D(R) &=& \int_R P(x|h) dx ,\\
P_F(R) &=& \int_R P(x|0) dx ,
\ea
respectively, where $R$ is the detection region (to be determined).

The Likelihood ratio $\Lambda$ is the ratio of the pdf when the signal
is present to the pdf when it is absent:
\begin{equation}\label{eq:likelihoodfunction}
\Lambda={P(x(t)|h(t)) \over P(x(t)|0)}.
\end{equation}

Taking the data to be $x(t)=h(t)+n(t)$, with $h(t)$ the signal and
$n(t)$ the noise and with the assumption that the noise is a
\textit{zero-mean, stationary} and \textit{Gaussian} random process,
we can write the likelihood ratio as
\ba \Lambda&=&{P(x|h)\over P(x|0)}\nonumber\\
&=&{\exp (-{1\over 2} (x-h|x-h))\over \exp (-{1 \over
2}(x|x))}\nonumber\\
&=&\exp[(x|h)-{1 \over 2}(h|h)]. \ea
This leads to the log of likelihood function as
\be \label{eq:log-likelihood} \log\Lambda = (x|h)-{1\over 2} (h|h).
\ee

We can also rewrite the simple expression of Eq.
(\ref{eq:log-likelihood}) for the likelihood function in terms of the
new variables, $A^a$ and $h_a$, as
\begin{equation}\label{eq:newloglikelihood}
\log \Lambda=(x|A^a h_a)-{1\over 2}(A^ah_a|A^bh_b).
\end{equation}
where the constant (in time) amplitudes $ A^a=A^a(h_0, \psi, i,
\Phi_0)$ are \cite{CS}
\begin{eqnarray}
  A^1 &=& ~~A_{+}\cos{\Phi_0}\cos{2\psi} -
  A_{\times}\sin{\Phi_0}\sin{2\psi},\label{eq:A1}\\
  A^2 &=& ~~A_{+}\cos{\Phi_0}\sin{2\psi} + A_{\times}\sin{\Phi_0} \cos{2\psi},\\
  A^3 &=& -A_{+}\sin{\Phi_0}\cos{2\psi} - A_{\times}\cos{\Phi_0} \sin{2\psi},\\
  A^4 &=& -A_{+}\sin{\Phi_0}\sin{2\psi} +
  A_{\times}\cos{\Phi_0}\cos{2\psi}\label{eq:A4}.
\end{eqnarray}
and
\begin{eqnarray}
h_1(t) &=& a(t)\cos \phi(t),\label{eq:h1}\\
h_2(t) &=& b(t)\cos \phi(t),\label{eq:h2}\\
h_3(t) &=& a(t)\sin \phi(t),\label{eq:h3}\\
h_4(t) &=& b(t)\sin \phi(t),\label{eq:h4}
\end{eqnarray}
where $a(t)$ and $b(t)$ are functions of right ascension $\alpha$
and declination $\delta$; they are independent of $\psi$, and $\phi$
is the phase of the wave signal seen at the Solar System Barycenter (SSB) \cite{jks}.
Likewise
\begin{eqnarray}
A_+ &=& {1\over 2}h_0 (1+ \cos^2{\iota}),\label{eq:aplus}\\
A_{\times} &=& h_0 \cos{\iota},\label{eq:across}
\end{eqnarray}
where $h_0$ is the wave amplitude, $\iota$ the inclination angle,
$\psi$ the polarization angle and $\Phi_0$ the initial phase.

Since the $A^a$s depend neither on the detector properties nor on the
frequency or the time, we can take them out of the inner product and
write the log of likelihood ratio as
\begin{equation}
\log \Lambda=A^a(x| h_a)-{1\over 2}A^aA^b(h_a|h_b).
\end{equation}
Defining the new variables
\begin{equation}\label{eq:H}
H_a\equiv(x| h_a),
\end{equation}
and
\begin{equation}\label{eq:M}
M_{ab}\equiv(h_a| h_b),
\end{equation}
we have
\begin{equation}
\log \Lambda=A^a H_a-{1\over 2}A^a A^b M_{ab}\label{eq:logLambda}.
\end{equation}
The maximum detection probability follows from the maximization of the
likelihood function: by maximizing the likelihood function with
respect to the $A^a$ (which, again, are independent of the detector),
we have
\begin{equation}
{\partial \log\Lambda\over \partial A^a}= 0.
\end{equation}
This leads us to
\begin{equation}
H_a-A^b_\text{\tiny MLE}M_{ab}=0 \label{Ha},
\end{equation}
and therefore
\begin{equation}\label{eq:Ab}
A^b_\text{\tiny MLE}= (M^{-1})^{ab}H_a.
\end{equation}
The label \textit{MLE} denotes the \textit{Maximum Likelihood
Estimator}; it corresponds to the values for the $A^a$s we calculate
from our data by maximizing the likelihood ratio (so that, in
practice, we are calculating $A^a= E[A^a_{\tiny MLE}]$). By
definition, the $\F$-Statistic is the \textit{maximum of the logarithm
of likelihood function}. Substituting Eq. (\ref{eq:Ab}) into Eq.
(\ref{eq:logLambda}), we have
\be\label{eq:fstat} \F\equiv \log \Lambda\mid_\text{\tiny
MLE}={1\over 2} ~H_a~ (M^{-1})^{ab}~H_b. \ee
This is the $\F$-Statistic which a generalized version of that for the
multi-IFO (Interferometer Observatory) can be found in
\cite{iraj-thesis}.

Again by writing the data as $x(t)=n(t)+h(t)$, and using Eq. (2.5) of
Cutler-Schutz (CS) \cite{CS} indicating that $\langle
(x|n)(y|n)\rangle=(x|y)$ with this fact that $\langle(h|n)\rangle=0$,
we would have the following
\be \langle2\F\rangle=4+(h|h),\label{eq:expectation-2F}\ee
which follows a $\chi^2$ distribution with 4 degrees of freedom and
non-centrality parameter $\rho^2\equiv (h|h)$, that is the square of
optimal signal to noise ratio (SNR$^2$). The degrees of freedom come
from the 4 \textit{unknown parameters of pulsar} namely the {\bf
amplitude ($h_0$), inclination angle ($\iota$), polarization angle
($\psi$)} and the {\bf initial phase ($\Phi_0$)}.  As was described in
\cite{CS}, even for a multi-IFO the number of freedom will remain
unchanged, since we always have the same 4 unknown parameters in
entire the search.

The $\F$-statistic shown above, which was originally derived by
Jaranowski, Kr\'olak and Schutz (JKS)\cite{jks}, is the optimal
statistic for detection of nearly periodic gravitational waves from GW
pulsars. We can use this statistical tool to search for any kind of
pulsars; unknown sources or targeted search. In targeted search we
know everything about the source by using some other astronomical
techniques, such as radio astronomy, gamma and X-ray astronomy
etc. The main information required for our search are; frequency and
its derivatives and position of the source. Given all these
information to the software developed by the LSC (LIGO Scientific
Collaboration) \cite{lal, lal-doxygen} (the implementation to our work
can be found in \cite{iraj-thesis}), and using a single workstation,
we will be able to search a known pulsar in a few minutes.

The work in this paper was done using simulated data for 100 arbitrary
pulsars. To set the positions and the frequencies of these simulated
pulsars, we have used the data of the known pulsars, as given in the
Australian Telescope National Facilities (ATNF) catalogue
\cite{ATNF}. To make the simulated data more realistic, we generated
the data at the level of LIGO detectors sensitivity. Since our
detectors (initial LIGO and even Enhanced LIGO) are not sensitive
enough to detect any signal until now, we assume that the data are
just simply noise without any signal in them. Due to that, our
simulated data are just pure noise and therefore instead of looking
for any detection, we set the upper limit on the strength of the
gravitational wave signal. For the historical reason we take the value
of 95$\%$ for the upper limit. All the search done in this paper are
in frequency domain. As the main goal of this paper is to compare two
different approaches in setting upper limits, we use Bayesian and
Frequentist algorithm to perform it on the same data for each
pulsar. These two algorithms will be explained in more details below.

\section{Frequentist Upper Limits}\label{sec:FreqUL}

The frequentist probability of an event represents the expected
frequency of occurrence of that event. The result for our upper limits
depends crucially on the experimental data under examination.  The
confidence value associated with these upper limits indicates the
expected occurrence of detection statistics values more significant
than the one that we have measured in the presence of signals whose
amplitude is equal to the upper limit value.

To set the frequentist upper limit on the amplitude of gravitational
waves, we use the $\F$-statistic as an optimal detection statistic. To
start with, we need to assign a confidence level $\mathcal{C}$ --
roughly speaking, our criterion will then be that, for our ``repeated
measurements'', in $\mathcal{C}$-percent of the time the value of
$2\F$ is above a specified threshold.

Let us explain how this works in detail for the example of setting a
95\% upper limit on $h_0$. For this, we need to find at which $h_0$ it
is true that 95\% of the values of the $\F$-statistic are above the
initial value of $2\F$ derived from the data.  To do so, we proceed step
by step as follows:

\begin{description}

\item{1.} Compute the $\F$-statistic of a perfectly matched signal
  using the exact values for the signal parameters (such as frequency,
  longitude, latitude and frequency derivatives). Let us call the
  resulting value of the $\F$-statistic $\F^*$.

\item{ 2.} Estimate the signal amplitude, $h_0$, using our parameter
estimation routine.

\item{ 3.} Take this $h_0$ as the initial value of the search.

\item{ 4.} Since we assume that there is no signal in the data -- that
  it is pure noise --, we can randomly assign arbitrary values to the
  other signal parameters (such as $\phi_0$, $\psi$ and
  $\cos{\iota}$).

\item{ 5.} To determine the probability distribution of
  $\F$-statistic, we take a random frequency value with a band of
  $0.1$Hz around the actual pulsar frequency (as was proposed in the
  LIGO S1 paper \cite{S1:pulsar}) and inject the artificial
  signal. With this choice, we are sure to be on the safe side; we use
  a large amount of data (in order of several months up to a year), so
  that the $0.1$Hz band will not lead to any spurious correlations
  between the search parameters.

\item{ 6.} After injection, compute the $\F$-statistic once again. Let
  us designate the resulting value of $2\F$ as $\F'$; store this value
  for later use.

\item{ 7.} Repeat the injections, computing of $\F$-statistic for 150
  times.  Save all resulting values of $\F'$. (The number of
  iterations used here is a heuristic value.)

\item{ 8.} As we are looking for a 95\% upper limit, proceed as
  follows: if the confidence level (the percentage of instances in
  which $\F'$ is greater than $\F^*$) was less than $90\%$ or above
  $98\%$ (say $x$), multiply the $h_0$ by the ratio of ${95\over x}$
  and take this value as the initial $h_0$ for the next step.

\item{ 9.} Repeat steps ``6'' and ``7'' until the confidence level is
  in one of the following ranges: a) $90\%-95\%$, or b) $95\%-98\%$.

\item{ 10.} For case a), multiply $h_0$ by 1.05; for case b), multiply
  by 0.90. (The factors 1.05 and 0.90 are, again, heuristic.)

\item{ 11.} Repeat the calculations of step ``7'' and following, but
  this time with 1000 injections in each run (instead of 150) to
  improve the statistics.

\item{ 12.} Repeat step ``11'' for 6 times; in each run follow the
  instructions in step ``10''. (The number of repetitions is
  heuristic; it is chosen in a way that the range of computed
  confidence levels will always include values higher and lower then
  95\%; therefore we can make an "interpolation" fit instead of having
  to extrapolate.)

\end{description}

\begin{figure}
  \begin{center}
    \includegraphics[width=\columnwidth]{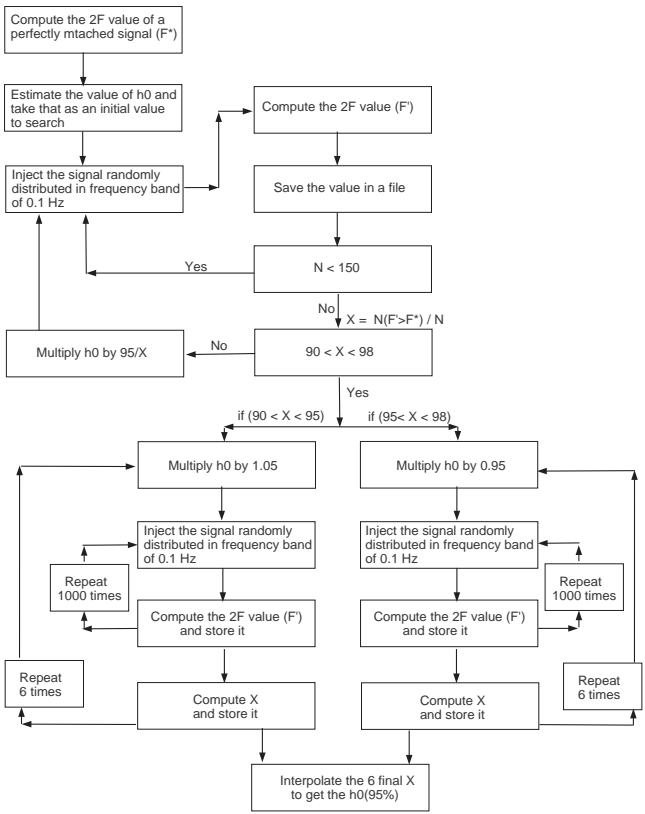}
    \caption{A Flow-Chart of how we implemented the frequentist upper
      limit.}
    \label{fig:Frequentist-UL-FlowChart}
  \end{center}
\end{figure}

A flow chart version of this procedure can be found in Fig.
\ref{fig:Frequentist-UL-FlowChart}.

\section{Bayesian Upper Limit}

The Bayesian probability is a measure of \textit{degree of belief} in
the occurrence of a statistical process. In contrast with the
Frequentist probability, in the Bayesian approach, we do not need for
an event of that particular type to have actually happened; all we
need is to find a measure for the degree to which a person believes
that a given proposition is true.

\subsection{Theoretical approaches}

The key ingredient of the Bayesian approach is the Bayes' theorem (a
simple proof of that can be found in \cite{cowan})
\be P(A|B)={P(B|A)P(A)\over P(B)}.\label{eq:bays-theorem}\ee
The term at the left hand side is called the \textit{posterior
probability}, while $P(A)$ is the \textit{prior probability} which
reflects our initial knowledge about the quantity $A$. The term
$P(B|A)$ is called the \textit{likelihood function}; the log of this
is, in fact, the $\F$-statistic to be computed from our data.

Our goal is to set an upper limit on the strength of the gravitational
wave amplitude, $h_0$, using a given amount of available data -- which
is the posterior probability of $h_0$ that we look for. Therefore, our
data plays the role of $B$ in Eq. (\ref{eq:bays-theorem}); which we
denote it by $s$. The term $A$ is the quantity about which we intend
to draw conclusions using our data; in our case, this is the upper
limit $h_0$, so we will substitute $h_0$ for $A$ in what follows.
With these substitutions, Eq.  (\ref{eq:bays-theorem}) now reads
\be P(h_0|s)={P(s|h_0)\times P(h_0) \over P(s)},\ee
where $ P(h_0|s)$ is the conditional probability of $h_0$ (posterior
probability) given the data $s$, $P(s|h_0)$ is the likelihood function
(to be defined below), and $P(h_0)$ represents our prior knowledge
about the distribution of $h_0$.

Since the term $P(s)$ is independent of our signal, we can consider it
as the constant normalization factor; it will cancel out automatically
when we compute the confidence level. Therefore, we can rewrite our
Bayes' theorem for the general case of all signal parameters as
\be P(h_0,\psi,\iota,\Phi_0|s)\propto P(s|h_0,\psi,\iota,\Phi_0)\times
P(h_0,\psi,\iota,\Phi_0), \ee
where, again, $P(h_0,\psi,\iota,\Phi_0|s)$ is our posterior
probability (to be calculated), $P(s|h_0,\psi,\iota,\Phi_0)$ is the
likelihood function and $P(h_0,\psi,\iota,\Phi_0)$ is the prior
probability of $h_0,\psi,\iota,\Phi_0$.

There are two common choices for estimating prior probability, known
as a \textit{flat prior} and \textit{Jeffrey's prior}. In the flat
prior, the prior probability is chosen to be constant ($P(h_0)\equiv
constant$), while in Jeffrey's prior, it is taken to vary inversely
proportional to the value of $h_0$ ($P(h_0)\equiv 1/h_0$). For more
details we refer the reader to \cite{gregory} and a comparison for
this case can be found in \cite{rejean-thesis}.

The Jeffrey's prior gives a higher value in upper limit than the flat
prior, while a flat prior gives a more realistic value for our case
\cite{rejean-thesis}. Therefore in the following we will focus on a
flat prior, as the case followed in \cite{iraj-thesis}.

To obtain the posterior probability, we need to calculate the
likelihood function. By Eq. (\ref{eq:fstat}), it can be expressed as
\begin{equation}
P(s|h_0,\psi,\iota,\Phi_0)\propto \text{e}^{-{1\over
2}M_{ab}(A^a-A^{a_0})(A^b-A^{b_0})}=G,
\end{equation}
where $A=(A^1,~ A^2,~ A^3,~ A^4)(h_0,\psi,\iota,\Phi_0)$ are the four
amplitude parameters defined in Eqs.  (\ref{eq:A1}-\ref{eq:A4}) and
$G=G(h_0,\psi,\iota,\Phi_0)$. The $A^0=(A^{1_0},~ A^{2_0},~ A^{3_0},~
A^{4_0})$ are also the best fit for the $A^a$s resulting from our
calculation of the $\mathcal{F}$-statistic.

For proper normalization, we first compute the integral
\begin{equation} \label{eq:I_FULL}
I\equiv \int_0^{\infty}P(h_0)dh_0 \int_{-1}^1 d\mu
\int_{-\pi/4}^{\pi/4} d\psi \int_0^{2\pi} d\Phi_0~
G,
\end{equation}
where $\mu \equiv \cos{\iota}$, and we will set `$P(h_0)\equiv
constant$'. To find the upper limit we use $h_0^{max}$ as the
upper bound in the integration over $h_0$,
\begin{equation}\label{eq:I_UL}
I_{UL}\equiv \int_0^{h_0^{max}}P(h_0)dh_0 \int_{-1}^1 d\mu
\int_{-\pi/4}^{\pi/4} d\psi \int_0^{2\pi} d\Phi_0~
G.
\end{equation}
We select $h_0^{max}$ in such a way that the ratio $I_{UL}/I$ gives us
the desired confidence level. In our case, we are looking for the
$95\%$ upper limit, therefore
\be\label{eq:I_ratio}
{I_{UL}\over I} = 0.95.
\ee
%

\subsection{Practical implementation}

To implement the above formalism, let us first construct the function
$G(h_0,\psi,\iota,\Phi_0)$. To do so, we need to expand the matrix of
Eq.  (\ref{eq:M}). The elements of this matrix depend on the three
amplitude modulation coefficients ($A,~B$ and $C$) defined in
\cite{CS}. Based on the notation used here, these elements take the
form of
\be M_{ab}=\left[
\begin{array}{cccc}
 A/2 & C/2 & 0   & 0\\
 C/2 & B/2 & 0   & 0\\
 0   & 0   & A/2 & C/2\\
 0   & 0   & C/2 & B/2
\end{array}
\right], \label{eq:M_ab}
\ee
which a detailed procedure of their derivation can be found
\cite{iraj-thesis}. Then we can construct the four elements
\ba
G_1 &=& {A \over 2}(A^1-A^{1_0})^2 + {C \over 2}(A^1-A^{1_0})(A^2-A^{2_0}),\\
G_2 &=& {B \over 2}(A^2-A^{2_0})^2 + {C \over 2}(A^2-A^{2_0})(A^1-A^{1_0}),\\
G_3 &=& {A \over 2}(A^3-A^{3_0})^2 + {C \over 2}(A^3-A^{3_0})(A^4-A^{4_0}),\\
G_4 &=& {B \over 2}(A^4-A^{4_0})^2 + {C \over
2}(A^4-A^{4_0})(A^3-A^{3_0}),
\ea
to make the final form of
$G(h_0,\psi,\iota,\Phi_0)$ in Eqs. (\ref{eq:I_FULL}) and
(\ref{eq:I_UL}) as
\be
G =\exp[-{1 \over 2}(G_1 + G_2 + G_3 + G_4)].
\ee
This is the core equation for our upper-limit analysis in Bayesian
approach. To construct this, we need all the above mentioned
parameters to be resulted from our software. The software we have used
for this purpose was developed partly by the author of this paper and
is now part of the LAL (LIGO Algorithm Library) \cite{lal}. With this
software we calculate the four amplitudes $A^a$ as well as the matrix
elements $M_{ab}$ (namely the amplitude modulation coefficients $A, B$
and $C$).  Once we have constructed the likelihood function
$G(h_0,\psi,\iota,\Phi_0)$, we can calculate the UL value in Eq.
(\ref{eq:I_ratio}) in two ways. One is to follow the exact procedure
spelled out above; first calculating the normalization in Eq.
(\ref{eq:I_FULL}) and then trying to find a value of $h_0^{max}$ for
which the ratio of Eq. (\ref{eq:I_ratio}) will be satisfied. This can
be done using the \textit{Numerical Integration} routines in
mathematical software like \textit{Mathematica} and
\textit{Maple}. Another way would be to calculate the posterior
probability of $h_0$ by marginalizing over the other three
parameters. This can be expressed in mathematical form as
\be p(h_0|s) \propto \int\int\int G(h_0,\psi,\iota,\Phi_0)~ d\psi~ d\mu
~d\Phi_0. \ee
Once the posterior probability for $h_0$ is known, one can then
integrate it over a sufficient range of $h_0$ to find out the area
covered; the result can be used for proper normalization (namely unit
total area). Next, we can find out at which $h_0$ the fraction of area
would satisfy our required confidence level.

Both the above methods have given equivalent results as discussed in
details in \cite{iraj-thesis}. However, for the work expressed in
this paper, we followed the second algorithm.

\section{Results and discussion}

\begin{figure}
  \begin{center}
    \includegraphics[width=\columnwidth]{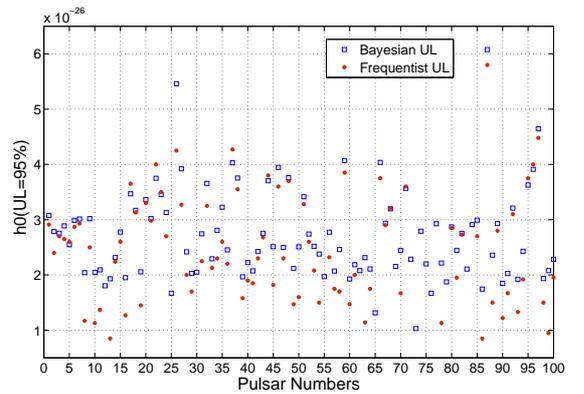}
    \caption[Upper Limits value of 100 pulsars.]{The upper limits
      value on the $h_0$ for 100 arbitrary pulsars using simulated
      data. In this plot, both Bayesian and Frequentist ULs are
      shown.}
    \label{fig:uls}
  \end{center}
\end{figure}

Once again, we have selected 100 arbitrary pulsars frequencies and
positions (based on the real pulsars information taken from ATNF
\cite{ATNF}). We have generated the simulated data at the level of
current LIGO detectors sensitivity (using the LAL software \cite{lal,
  lal-doxygen} developed by LSC) and computed the UL for these
pulsars. To compare with the real data, the simulated data contains
just noises where the upper limit set on them are shown in
Fig. \ref{fig:uls}.

The blue rectangular in this plot represent the value of upper limits
in Bayesian approach and the red circle points to the Frequentist
ones. The horizontal axis indicates the pulsars number, therefore on
each vertical line corresponds to each pulsar we should have one blue
rectangular and one red circle. However, as seen, in some cases there
is just one blue rectangular and missing red circle (Frequentist
UL). These count for 16 pulsars, in which 7 of them were caused due to
some unknown reason. The reason for the 9 others is that; in these
cases, due to very small amount of $2\F$, the Frequentist upper limit
procedure could not be converged to any particular value. It means,
for some cases, by changing (increasing or decreasing) the value of
$h_0$ for more than two order of magnitude, the upper limit value
always stands above 96\% or 97\%.  

\begin{figure}
  \begin{center}
    \includegraphics[width=\columnwidth]{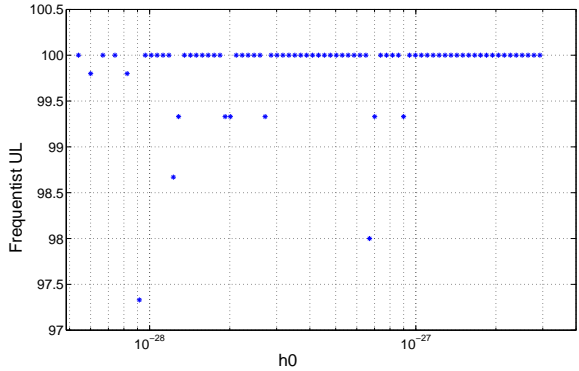}
    \caption[Bad Frequentist upper limits.]{One of the bad upper limit
      value on the $h_0$ in Frequentist approach. The value of $2\F$
      for this case was $0.08$.}
    \label{fig:bad_ul}
  \end{center}
\end{figure}

An example of that is shown in Fig. \ref{fig:bad_ul}, that is the case
where $2\F=0.08$. This figure shows the dependency of Frequentist
upper limit to $h_0$ for an individual pulsar. It's clear that even by
increasing the $h_0$ for about $2.5$ order of magnitude, the value of
upper limit lies mostly about 100\%. While, in general the upper limit
is very sensitive to small changes in $h_0$. This was done 76 times
and in each time the value of $h_0$ was increased according to the
procedure expressed in Sec. \ref{sec:FreqUL}. Note that the starting
and ending value of $h_0$ are significantly smaller than the $h_0$
require for $95\%$ upper limit shown in Fig. \ref{fig:uls}. Means
that, in normal condition where the required $h_0$ to get $95\%$ upper
limit is in order of $10^{-26}$, by setting the $h_0$ in the range of
$10^{-28}-10^{-27}$, we should get a very small upper limit compare to
$95\%$. In fact, as disscused below, the low value of $2\F$ for this
pulsar is the reason of such a behavior in the Frequentist framework.

The same behavior was shown in \cite{iraj-thesis, iraj-defense} by
using the real data. The reason is clear; we have pointed out that the
Frequentist approach is based on the number of occurance of an
event. For this we should set a threshold and count how many times the
value of that particular parameter is passing this
threshold. Naturally there can be some False Alarms (FA), which a
noise shows itself strong enough to pass this threshold. Since the
$\F$-statistic follows a $\chi^2$ distribution with four degrees of
freedom, the FA follows as (equation 3.44 of \cite{iraj-thesis}),
\be \alpha = (1+2\F)~\mbox{e}^{-2\F}. \ee

The values of $2\F$ in which the Frequentist upper limit could not be
converged are: $0.08, 0.34$, $0.46, 0.47$, $0.49, 0.51$, $0.59, 0.63$
and $0.84$.  So, by putting these numbers in the above equation we get a
very large values of FA. For example, in the case of $2\F=0.84$
we have $\alpha=0.80$, $2\F=0.46$ gives $\alpha=0.92$ and for
$2\F=0.08$ we get $\alpha=0.997\sim 100$. This explains the whole
story.

This high false alarm probability means that {\it noise alone} has a
high chance of producing an $\F$-statistic value greater than the
$\F^*$ produced by our data set and our template. In other words, our
realization of the noise is one that; it is particularly {\it
  unlikely} to look like it contains our signal, and this statistical
fluctuation yields a low upper limit value or even does not converge
-- due to the nature of the procedure that we use to determine the
upper limit. The Bayesian approach is less sensitive to such
fluctuations. Note that the resulting frequentist upper limit is {\it
  not} an artifact of our technique, but it is still a perfectly
correct and consistent upper limit in the Frequentist framework. This
clearly shows the nature of our data.

\begin{figure}
  \begin{center}
    \includegraphics[width=\columnwidth]{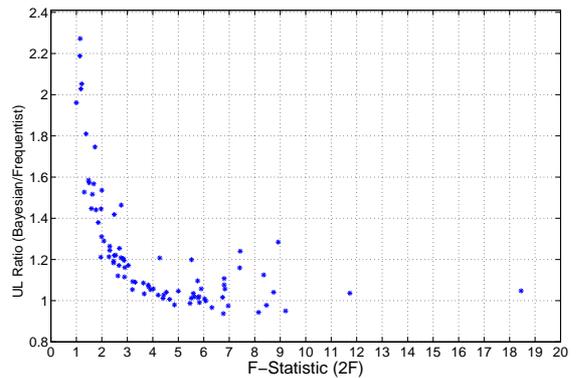}
    \caption[Upper Limits ratio of 100 pulsars.]{The ratio of Bayesian
      over Frequentist upper limits value on the $h_0$ for 100
      arbitrary pulsars using simulated data.}
    \label{fig:ul_ratio}
  \end{center}
\end{figure}

To investigate further and check the behavior of the Frequentist UL,
let us compare its value with that of Bayesian. To do so, we plot the
ratio of the Bayesian UL over the corresponding value of Frequentist
for the same pulsar versus the value of $2\F$ (Fig.
\ref{fig:ul_ratio}). The output says; as we go further to the value of
$2\F$ less than $4$, the ratio increases and when we reach to the
$2\F=1$, this ratio is quite significant; about a factor of $2.3$. For
the case of $2\F<1$, we have already seen that the upper limit for the
Frequentist approach did not converge. Therefore they are not shown in
this plot. If they would converge, they should be quite lower than
that of Bayesian and therefore the ratio should go much higher. The
same behavior was shown in \cite{iraj-thesis} with real data. 

Fig. \ref{fig:ul_ratio} also shows that, at roughly $2\F=4$ the ratio
is close to unity and roughly remains the same when $2\F>4$. This
tells that the problem of low value in upper limit in Frequentist
approach appears when we have $2\F<4$ while for larger value of
$\F$-statistic there is always agreement between Frequentist and
Bayesian frameworks.

Apart from the difference in the nature of Frequentist and Bayesian
frameworks, there is another difference in performing a Frequentist
and the Bayesian upper limit search. Since in Frequentist algorithm we
need to inject some artificial signals into the data and then search
the newly generated data to compute the $\F$-statistic in each
iteration, this requires a high amount of computational resources. To
increase the sensitivity we need to use more data that requires more
computational power as well. Because, the required time to search the
data to compute the $\F$-statistic is linearly proportional to the
amount of data. In order to have a better statistic in Frequentist
algorithm, we therefore need a larger iteration. This would
additionally brings another linear increment in the cost for the
computation.  While in the Bayesian approach, to compute the
$\F$-statistic and the other components, we search the data just
once. Then compute the $P(h_0)$ by marginalizing the probability over
the $\psi, \iota$ and $\Phi_0$. These all will be done once and are
computationally very cheap. As an estimation, the entire process for
one pulsar using Bayesian algorithm takes about half an hour up to one
hour in a single workstation. In a good approximation this is
independent of the amount of data. Because, searching in the large
amount of data (say about one year) to compute the $\F$-statistic and
other components takes just about few minutes. In contrary, the
required time for a Frequentist algorithm to search for single pulsar
in an amount of data in order of one year takes about \textit{3 weeks}
on a single workstation.

As a summary; although search in the Frequentist upper limit shows the
exact nature of our data, however there are some disadvantages with
the same search by using the Bayesian algorithm. The important one is
that in the case where our data shows a small value of $\F$-statistic
in a particular frequency bin and position of the pulsar, we cannot
trust the upper limit value produced by Frequentist
approach. Likewise, performing a search in the Frequentist framework
is much expensive than the same search in Bayesian approach. 

\section{Distribution of 95\% Bayesian upper limits on $h_0$
  using simulated data in frequency domain}

As an application of Bayesian algorithm, we now present the
distribution of ULs (95\% upper limit on $h_0$) computed in the
Frequency Domain (FD). We start with the idealized case of a large number
(5500) of simulated data set with pure noises (no signal), and compute
the 95\% upper limit on $h_0$ of each data set. The sky locations in
the search are chosen randomly such that their distribution over the
solid angle is uniform; detectors position are picked randomly from a
list consisting of the locations of H1, L1, VIRGO and GEO600
detectors. The resulting mean upper limit is
\be \langle h_0^{95\%}\rangle = (10.67 \pm 0.04) \sqrt{S_h(f) \over
T}. \label{eq:average-distribution-h95} \ee

In order to compare our result with the simulation in Time Domain (TD)
done by Dupuis and Woan \cite{rejean-graham}, we repeated this
experiment with only H1, L1 and GEO600 detectors, as in their
analysis. The results are in a very good agreement with a ratio in ULs
\be
{\langle h_0^{95\%}\rangle_{FD} \over \langle h_0^{95\%}\rangle_{TD}} = 0.98.
\ee

\begin{figure}[h]
  \begin{center}
    \includegraphics[width=\columnwidth]{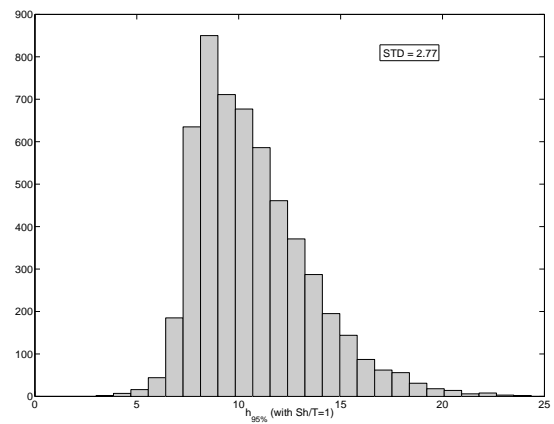}
    \caption[Distribution of 95\% upper limit on $h_0$ for simulated
    data.]{Distribution of 95\% Bayesian upper limit on $h_0$ using 5500 
      individual simulated data runs in frequency domain. $\langle
      h_0^{95\%}\rangle = (10.59 \pm 0.04)$.}
    \label{fig:h95-disttributions}
  \end{center}
\end{figure}

Fig. \ref{fig:h95-disttributions} shows the distribution of
$h_0^{95\%}$ for 5500 different runs in frequency domain, which is also
in a good agreement with that of presented in
\cite{rejean-graham}. 

These results show that although we use different domain (TD or FD) to
search for gravitational waves, if we stay in Bayesian framework, both
give the same results theoretically (a more details can be found in
\cite{iraj-thesis}). However, using different frameworks (Frequentist
or Bayesian) will may lead to a different outcome.
%


\end{document}